\documentclass[aps,twocolumn,groupedaddress,showpacs]{revtex4}
\usepackage{graphicx}
\def\pderiv#1#2{\frac{\partial #1}{\partial #2}}
\def\r{{\bf r}}

\def\sci#1#2{#1\times 10^{#2}}
\def\gtrsim{\mathrel{\rlap{\raise .5ex\hbox{$\textstyle >$}}
  {\lower .7ex\hbox{$\sim$}}}}
\def\lesssim{\mathrel{\rlap{\raise .5ex\hbox{$\textstyle <$}}
  {\lower .7ex\hbox{$\sim$}}}}
\def\Prob#1{{\rm Prob}\left\{#1\right\}}
\def\const{{\rm const}}
\def\Re{{\rm Re}}
\def\Im{{\rm Im}}
\def\mod{\,{\rm mod}\,}
\def\erf#1{{\rm erf}\left(#1\right)}
\def\erfc#1{{\rm erfc}\left(#1\right)}
\def\erfinv#1{{\rm erf}^{-1}\left(#1\right)}

\def\efficiency{{\cal{E}}}

\begin{document}
\bibliographystyle{apsrev}

\title[Acoustic Detection of Neutrinos]{Sensitivity of an
underwater acoustic array \\
to ultra-high energy neutrinos}

\author{Nikolai G. Lehtinen}
\author{Shaffique Adam}
\thanks{Now at Cornell University}
\author{Giorgio Gratta}
\affiliation{Physics Department, Stanford University, Stanford, CA}
\author{Thomas K. Berger}
\author{Michael J. Buckingham}
\affiliation{Scripps Institution of Oceanography, UC San Diego, La Jolla, CA}

\date{\today}

\begin{abstract}
We investigate the possibility of searching for ultra high energy
neutrinos in cosmic rays using acoustic techniques in ocean water.
The type of information provided by the acoustic detection is
complementary to that of other techniques, and the filtering effect of
the atmosphere, imposed by the fact that detection only happens if a
shower fully develops in water, would provide a clear neutrino
identification.  We find that it may be possible to implement this
technique with very limited resources using existing high frequency
underwater hydrophone arrays.  We review the expected acoustic signals
produced by neutrino-induced showers in water and develop an optimal
filtering algorithm able to suppress statistical noise.  The algorithm
found is computationally appropriate to be used as a trigger for the 
signal processors available on existing arrays.  We estimate the noise
rates for a trigger system on a very
large size hydrophone array of the US Navy and find that, while a
higher density of hydrophones would be desirable, the existing system
may already provide useful data.
\end{abstract}
\maketitle

\section{Introduction}
\label{sec:introduction}

The understanding of the highest energy cosmic rays represents one of
the most challenging fields of modern physics.  While abundant high
quality data has greatly enhanced our knowledge of cosmic rays of
energies up to several tens of GeV, the study of higher energies is
limited by the low fluxes available. Yet the study of particles with
energies in excess of $10^{18}$~eV (ultra-high energy cosmic rays, or
UHECR) promises boundless opportunities of discovery.  To date some
12 cosmic ray events have been observed with energy in excess of
$10^{20}$~eV~\cite{watson00}.  While the acceleration mechanisms at
these energies are not completely understood~\cite{olinto00}, even more
fundamental problems arise from the apparent inconsistency of data
with the Greisen-Zatsepin-Kuzmin (GZK) cutoff. Such
cutoff~\cite{greisen66,zatsepin66} is expected to limit the maximum
energy of protons of cosmological origin somewhere below $10^{20}$~eV,
because of the finite ($\approx 50$~Mpc) inelastic collision length of
such particles in the cosmic microwave background radiation (CMBR).
Indeed the inelastic scattering of UHE protons off CMBR should at the
same time suppress the proton flux and breed neutrinos from decay-products
of pions.  While the present data seem to indicate that photons and
neutrinos are not the main component of UHECR~\cite{watson00},
reliable identification of the primary particle type and its energy
are essential parameters for the study of this problem.

It was recently pointed out~\cite{halzen00} that a substantial UHE
neutrino component may accompany UHE protons and nuclei, due to
neutrino production in cosmic beam dumps.  Neutrino production at
ultra-high energies appears in fireball models of gamma-ray
bursts~\cite{vietri98,waxman95,waxman97}, active galactic
nuclei~\cite{gaisser95,protheroe96a,protheroe96b,protheroe98,
mannheim95,stecker96} and, to a lesser degree, in Galactic
mechanisms~\cite{domokos93} and, as mentioned, the GZK
process~\cite{yoshida93}.  Another hypothetical source is the decay of
heavy objects predicted by some theories, known as the ``top-down''
models of UHECR production~\cite{olinto00,sigl97,sigl99}.

Weakly interacting neutrinos could, unlike UHE gamma rays and protons, 
reach us from distant and powerful sources, opening a deeper horizon for 
astrophysics, cosmology and, possibly, high-energy particle 
physics~\cite{gaisser95,gandhi96}.    While atmospheric neutrinos
represent an irreducible background for Earth-based detectors, such
background is expected to be modest because of the extremely long
decay length of pions at the energies of interest.

Three techniques have been used until now to detect UHE cosmic-rays, 
all involving the showering of the primary 
particle in the Earth's atmosphere.  The shower is then detected 
either by observing the fluorescence or \v{C}erenkov light induced 
by the ionizing tracks in the air, or by directly detecting the 
charged particles in the shower tail with scintillation counters 
scattered on the ground~\cite{ohalloran98,weekes96,watson00}.  
While some of the these techniques provide the largest acceptances
obtained in particle detectors, in general the study of UHECR is still
hampered by the very low flux one has to be sensitive to.  Typical 
fluxes are $\sim100$~km$^{-2}$y$^{-1}$ above $10^{18}$~eV, 
$\sim1$~km$^{-2}$y$^{-1}$ above $10^{19}$~eV, and
$\sim1$~km$^{-2}$century$^{-1}$ above $10^{20}$~eV~\cite{watson00}.
Dedicated neutrino telescopes using \v{C}erenkov light under water 
and the Antarctic ice-cap are for the time being optimized for the 
TeV to EeV energy-region~\cite{halzen00}.

While much has been learned from the above detectors, it is important
to explore, in parallel, new methods that could either increase the
flux sensitivity, and hence raise the energy threshold for detection,
or help constraining the primary particle identification and its
energy.  Alternative methods being discussed~\cite{gaisser95} include
the detection of radio \v{C}erenkov emission from the lunar
soil~\cite{gorham99} (sensitive only to neutrinos that can cross the
moon and interact upon exiting the satellite from its near side) and
active radar detection of showers in the atmosphere~\cite{gorham00}
(which has different systematics for particle identification since it
is more sensitive to horizontal showers than other techniques).  As
discussed in this paper, another possibility consists in the detection
of showers by means of the acoustic energy released in the medium
where they develop.  While, at least in principle, this technique
could use as a radiator either the soil of the moon or a large body of
water on the earth, here we concentrate on this second case that is
more practical and, as we will show, may be possible to test on a
large scale with a very modest effort.  It is important to realize
that these two radiators would provide very different information, the
first being sensitive to UHECR of any type, while the second being
sensitive essentially only to neutrinos because of the filtering
effect of the earth's atmosphere.  While the idea of taking advantage
of the very high mechanical $Q$ for small oscillation amplitudes and
ultra-low seismic noise characteristic of the moon was first
mentioned in~\cite{learned89}, acoustic detection of ionizing
particles in water was proposed in~\cite{askarjan57} and then
developed~\cite{learned79} in connection with the DUMAND
project~\cite{bradner77,parvulesku80,roberts92}. Although the primary
goal of DUMAND was optical \v{C}erenkov detection of muon tracks in
deep ocean water, the hydrophones, originally conceived to monitor
photomultiplier positions, were proposed to be used for acoustic 
detection of neutrinos of $>\!10^{16}$~eV energy. These early
ideas were nicely complemented by experimental data collected at
accelerators~\cite{sulak79}.  We will review the past work on the
subject and present a model useful to study UHE neutrino detection in
sea water.  We will then use this model to simulate signals and
develop an algorithm to optimally filter hydrophone data to extract
the expected signals from statistical noise.  While most the
assumptions for the study are rather general, we will apply our
results to the case of a large, high frequency test array that the US
Navy operates off the coast of Florida.

We note here that UHE neutrino fluxes substantially
larger than the UHECR flux quoted by the experiments utilizing air
showers could have gone unobserved.
The neutrino cross section due to neutral and charged current
interactions in the UHE regime is calculated in~\cite{kwiecinski98}.
Even for $E_{\nu}\sim 10^{20}$~eV the probability of interaction in
traversing entire atmosphere's depth is only of the order of $10^{-5}$.
Many models~\cite{protheroe96a,
protheroe96b,protheroe98,mannheim95,stecker96,waxman97,sigl97,sigl99},
predict neutrino fluxes higher than the flux of the UHECR observed in the
atmosphere.   For example, fireball gamma-ray burst
models predict a neutrino flux at energies $E_\nu>10^{20}$~eV as high as
$\sim$1~km$^{-2}$y$^{-1}$, active galactic nucleus models
$\sim\!\!0.1$~km$^{-2}$y$^{-1}$, and top-down models
$\sim\!\!10$~km$^{-2}$y$^{-1}$, as quoted in
\cite{kwiecinski98}.

\begin{figure}
\includegraphics{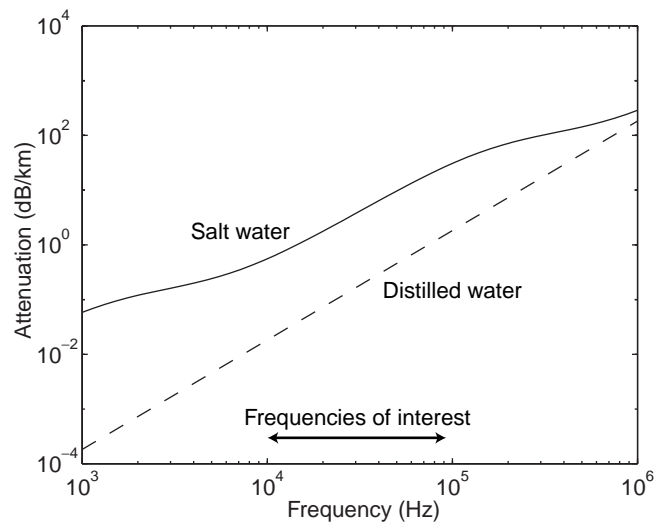}
\caption{Sound attenuation coefficient at 25$^\circ$C, in the sea and
distilled water as function of frequency~\cite{fisher77}.}
\label{fig:attenuation}
\end{figure}

\begin{figure}
\includegraphics{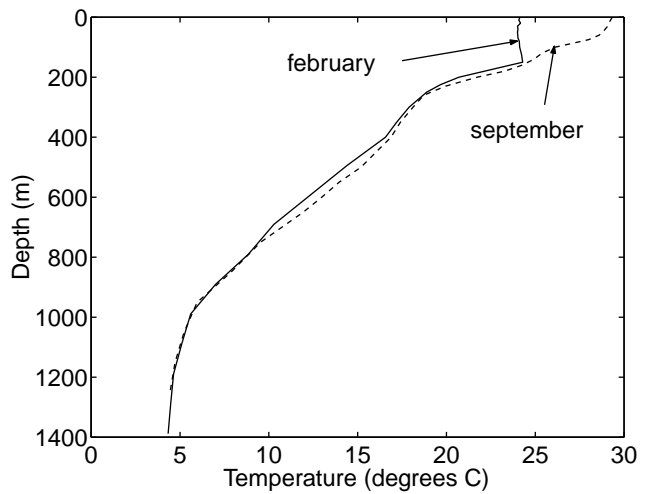}
\caption{Water temperature profiles for the tropical region of
of the AUTEC array discussed below~\cite{autec94}. The two curves 
correspond to the extreme variations of temperature through the year.}
\label{fig:temperature}
\end{figure}

\section{Acoustic signature of particles in water}
\label{sec:acoustic}

At the energies of interest here, neutrinos interact in water by a deep
inelastic scattering on quarks inside oxygen and hydrogen nuclei.  The
scattering produces a lepton and an hadronic shower, with similar energies
shared among these two components.
The structure of hadronic showers is the same for all three flavors of
(anti)neutrino. The behavior of the leptons, however, is different.
In the case of $\nu_{\rm e}$'s the lepton energy goes in an
electromagnetic shower and is essentially detected together with the
hadronic energy. For $\nu_{\mu}$'s calculations by
Mitsui~\cite{mitsui92} show that the mean-free-path in water between
catastrophic bremsstrahlung and direct pair production events with
energy transfer greater than $10^{19}$~eV is large (several km) with
respect to the vertical size of the detectors considered here.  The
residual ionization along the track is low, so that muons are
virtually undetectable by acoustic methods~\cite{parvulesku80}.  Tau
neutrinos create $\tau$ leptons, whose mean free path turns out to be
long enough to leave the detector volume for
$E_{\nu_\tau}\gtrsim10^{17}$~eV.

\begin{figure*}
\includegraphics{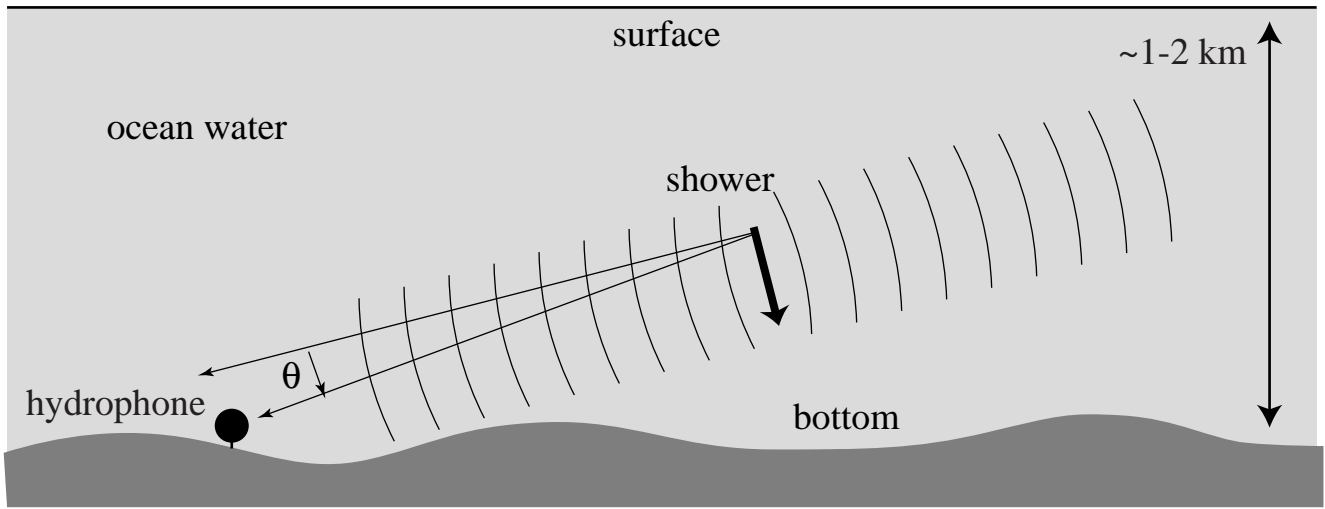}
\caption{Geometrical configuration of a shower and acoustic pulse produced. 
The angle $\theta$ is calculated with respect to the forward direction
from the perpendicular to the shower axis. The origin is at the
starting point of the shower. }
\label{fig:config}
\end{figure*}

At $E_{\nu}\sim 10^{20}$~eV in water a hadronic shower deposits 90\% of its
energy in a cylinder of some 20~cm radius and 20~m length. Sound is
produced in water mainly by heating localized along the shower, resulting in
volume expansion, as first suggested in~\cite{askarjan57}.  The shower
development (and hence the energy deposition) occurs at the velocity
of light and can be regarded as instantaneous for the purpose of
acoustic phenomena.  As already mentioned the sound generation has
been confirmed experimentally using artificial particles~\cite{sulak79}.  
Although it is not completely clear to what
extent these accelerator experiments, simulating 
$\sim$10$^{20}$~eV energies using bunches of over 10$^{11}$ protons
of $\sim$200~MeV, can describe the details of the lower energy-density
UHE neutrino interactions, their accuracy is probably sufficient for
the present study.

We analyze the acoustic signal production following~\cite{learned79}.
Let the energy deposited per unit volume per unit time be given by a function 
$E(\r,t)$.    The total neutrino energy is $E_0=\int_VE(\r)\,d^3\r$.
The wave equation for the pressure pulse produced $p$ is:
\begin{equation}
\nabla^2\left(p+\frac{1}{\omega_0}\dot{p}\right)-\frac{1}{c^2}\ddot{p}=
-\frac{\beta}{C_p}\pderiv{E}{t}
\label{eq:p}
\end{equation}
Where we use the parameters for sea water: $c\approx1500$~m/s is the
speed of sound, $\beta\approx\sci{1.2}{-3}$~K$^{-1}$ is the bulk
coefficient of thermal expansion,
$C_p\approx\sci{3.8}{3}$~J~kg$^{-1}$K$^{-1}$ is the specific heat at
constant pressure, and $\omega_0\approx\sci{2.5}{10}$~s$^{-1}$ is the
characteristic attenuation frequency. $\omega_0$ is,
strictly speaking, a function of frequency~\cite{fisher77}, as can be
seen from the plot of the attenuation coefficient $\alpha({\rm
dB/km})=(10^4/\ln 10 \omega_0 c)(2\pi f)^2$ in
Figure~\ref{fig:attenuation}.  For simplicity of calculations, we
assume that $\omega_0$ is a constant in the frequency range
$f=10-100$~kHz characteristic of the signal.

We note here that the coefficient of thermal expansion $\beta$ depends
upon the water temperature. It vanishes at $\sim\!-3^\circ$C for
typical sea water of 3.5\% salinity~\cite{crc}.  In the case of
vanishing $\beta$ other mechanisms of energy coupling to acoustic
modes have been proposed~\cite{hunter81,lyamshev92}.
Extreme temperature
profiles as a function of depth are given in
Figure~\ref{fig:temperature} for the tropical waters
of the site discussed below. As we can see, although
$\beta$ decreases with depth, it does not
reach zero, so we will restrict ourselves to the case of thermal
emission mechanism.

The instantaneous nature of the heating mechanism can be expressed by
$E(\r',t)= E(\r')\delta(t)$.  The pressure wave can then be calculated
at location $\r$ as a function of time $t$ as
\[
p(\r,t)=\int_V E(\r')G(\r-\r',t)\,d^3\r'
\]
where $G(\r,t)$ is the pressure pulse generated by a point source
$E=\delta(\r)\delta(t)$, taking attenuation into account:
\begin{equation}
G(\r,t)=-\frac{\beta}{4\pi C_p}
\frac{(t-r/c)}{r\sqrt{2\pi}\tau^3}e^{-(t-r/c)^2/(2\tau^2)}
\label{eq:Gatt}
\end{equation}
where $\tau=\sqrt{r/(\omega_0c)}$. 

The region of the energy deposition is elongated in the direction of the 
initial velocity of the neutrino, which constitutes the axis of the cascade. 
The acoustic emission is coherent in the plane perpendicular to this axis. 
Thus, the radiation diagram has a pancake shape, perpendicular to the shower
axis, as shown in Figure~\ref{fig:config}. This effect was also
observed in the 30~cm long and 4.5~cm wide cylinder energy deposition
by particles in one of the accelerator experiments described in~\cite{sulak79}.

We calculate the time dependence of the pressure at a distance
$r=1$~km from the origin, for different directions of observation. As
shown in Figure~\ref{fig:config}, the angle $\theta$ is calculated
in the forward direction from the perpendicular to the shower axis and
the origin is at the starting point of the shower. Since the maximum
energy deposition does not occur at the beginning of the shower, the 
maximum acoustic pulse results at some angle $\theta>0$.

\begin{figure*}
\includegraphics{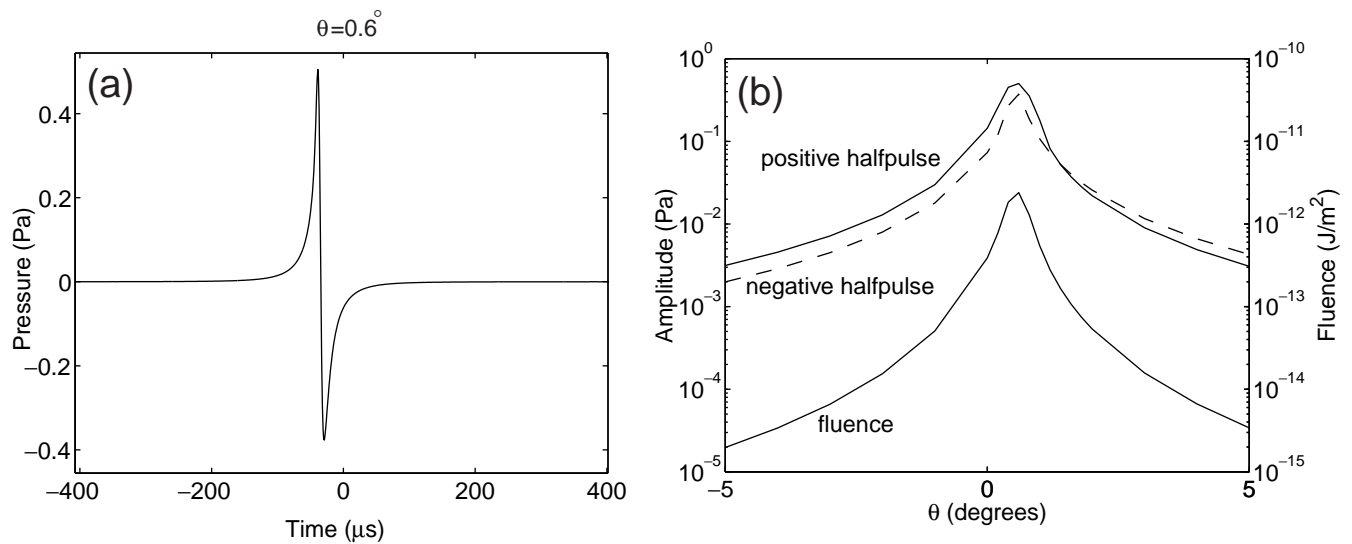}
\caption{Results of calculations of the acoustic signal from the
hadronic part of the neutrino-induced shower~\cite{alvarez98}, at the
distance of 1000~m from the shower axis, for a primary hadronic energy of 
$10^{20}$~eV: (a)~the pulse shape at the observation
angle of $\theta=0.6^\circ$, where the amplitude is maximal; (b)~the
pressure amplitude of the pulse and the total energy fluence in 
the pulse, $(\rho c)^{-1}\int_{-\infty}^{+\infty}p^2\,dt$. These last
two quantities are plotted as functions of observation angle as 
defined in Figure~\ref{fig:config}. }
\label{fig:alvhad}
\end{figure*}

To calculate the effect of a $10^{20}$~eV hadronic shower, we use a
model based on the results of one-dimensional Monte Carlo simulations
in ice~\cite{alvarez98}. This model includes the interactions of
$\pi^0$ and other short-lived resonances, as well as the
Landau-Pomeranchuk-Migdal (LPM)
effect~\cite{landau52,landau53,migdal56,migdal57}, that are important
at UHE. The transverse structure is modeled
after~\cite{sokolsky89}. The results of acoustic emission by a
hadronic shower are presented in Figure~\ref{fig:alvhad}.

The pressure pulse has a typical bipolar shape and is shown, at the
maximum of the radiation pattern, in Figure~\ref{fig:alvhad}a.  The
peak pressure and energy fluence for other angles of observation are
shown in Figure~\ref{fig:alvhad}b.  The signal at other observation
angles has a smaller amplitude and is stretched in time, still
preserving the bipolar shape.

\begin{figure*}
\includegraphics{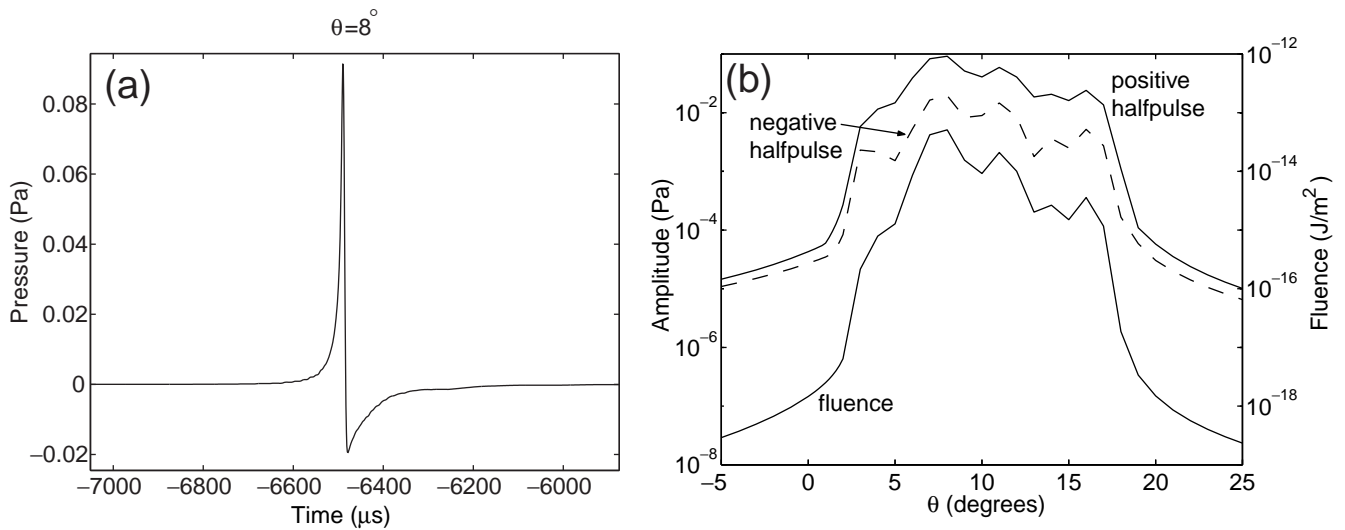}
\caption{The results of calculation of the acoustic signal from the
electromagnetic part of a $\nu_e(\bar{\nu_e})$-induced
shower~\cite{alvarez97,alvarez98}, at the distance of 1000~m from the 
shower axis. The total energy in the electromagnetic shower is
$10^{20}$~eV.  (a)~the pulse shape at the observation angle of
$\theta=8^\circ$, where the amplitude is maximal; (b)~the amplitude of
the pulse and the total energy fluence in the acoustic pulse, $(\rho
c)^{-1}\int_{-\infty}^{+\infty}p^2\,dt$. Both are plotted as functions
of observation angle as defined in Figure~\ref{fig:config}.}
\label{fig:lpm}
\end{figure*}

To calculate the case of an electromagnetic shower we use the Monte 
Carlo model {\em LPMSHOWER}~\cite{alvarez97} that includes the LPM effect.
In this case, with all the energy in the electromagnetic channel, the 
LPM effect dominates the shower shape, resulting in very large fluctuations,
and a significant elongation for initial energies $\gtrsim$1~EeV. 
For $E_e=10^{20}$~eV the shower is $\sim$300~m long.
The non-uniform energy deposition is the result of individual sub-showers 
starting at the locations of large energy loss. These sub-showers also create 
peaks in the acoustical radiation pattern. The properties of a typical 
shower are presented in Figure~\ref{fig:lpm}. The multi-peak 
structure is clearly visible in the amplitude and fluence diagrams as 
function of the angle.  While the pressure pulse shape, taken at the 
direction of maximum emission, has, also in this case, a simple bipolar 
shape, the maximum occurs at a large angle respect to the previous case, 
because of the longer shower profile.

The signal from electron (anti)neutrinos is the superposition of these
two cases in a proportion corresponding to the way the energy of the
primary is shared between the hadronic and electromagnetic shower.  
It was found~\cite{gandhi96} that at $E_{\nu(\bar{\nu})}=10^{20}$~eV
the hadronic component accounts for $\sim$20\% of the total energy.
For an UHE primary, the axes of the hadronic and electromagnetic showers 
are practically parallel, the angle between them being of the order of 
$10^{-6}$ radians~\cite{mitsui92}.

We note here that the peak pressures predicted at 1000~m from
$10^{20}$~eV showers are well within the sensitivity of good quality
hydrophones.  Indeed typical sensitivities for the frequency band of
interest are $\sim\!10^{-3}$~Pa~\cite{sulak79}.  However it is rather
clear that the two factors limiting the power of this technique will
be the ambient noise and the characteristic emission pattern described
above.  Such pattern substantially limits the solid angle accessible
to each sensor and hence, together with the noise level, it will
dictate the maximum tolerable spacing between detection sites.

\section{The AUTEC array as a UHE neutrino detector}
\label{sec:implementation}

\begin{figure}
\includegraphics[scale=0.45]{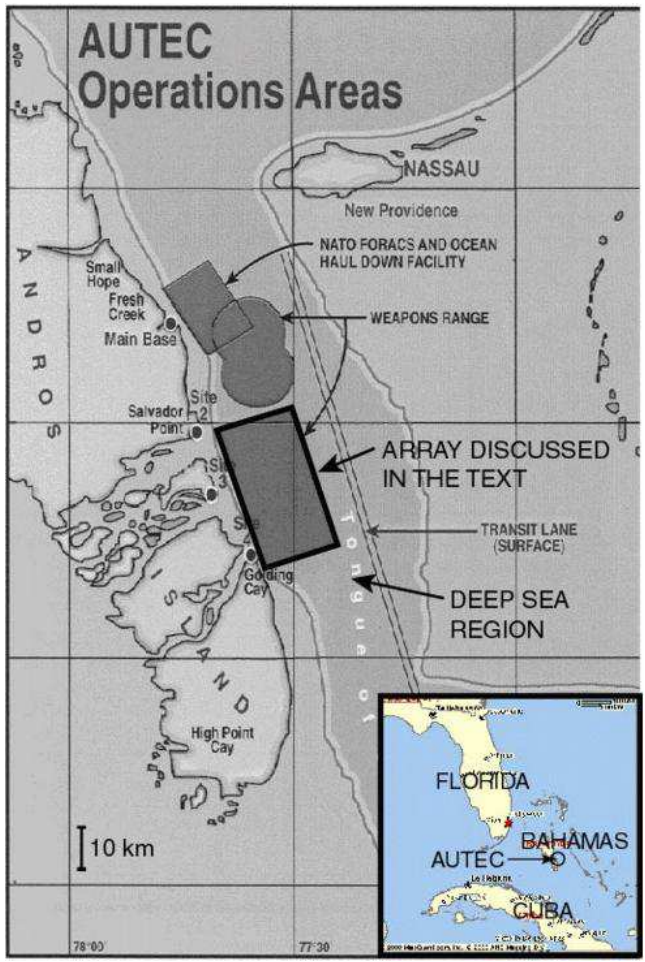}
\caption{Schematic view of the AUTEC facilities. The hydrophone array
discussed in the text is the larger quadrangular region.}
\label{fig:autec}
\end{figure}

While the installation of hydrophones to complement underwater
\v{C}erenkov arrays has been discussed by several
groups~\cite{domogatsky00,butkevich98}, it is interesting to
consider whether a very large existing array could be used in parasitic
mode for UHE neutrino detection.  Although the scientific community
has recently started discussing the use of long-range, early warning
military arrays for a variety of oceanographic
purposes~\cite{sosus},
the low bandwidth of
these systems, essentially designed to detect low-frequency ship noise
that can propagate over very large distances, make them quite
unsuitable for the type of signals discussed here.  Another type of
arrays exists, generally designed to track ships and weapons equipped
with special high frequency ``pingers'' during limited-range naval
exercises.  Such arrays have a relatively high-density of hydrophones 
with typical bandwidths in the tens of kHz.  The array considered here, 
at the Sites 3 and 4 of the Atlantic Undersea Test and Evaluation Center 
of the US Navy (AUTEC)~\cite{autec}, covers an approximate area of 250~km$^2$
($\approx 5\times 15$~nautical miles$^2$) with depths between 1400 and
1600~m, as shown in Figure~\ref{fig:autec}.

AUTEC is located in the ``Tongue of the Ocean'', a $50\times 200$~km$^2$
tract of deep sea, bounded to the west by Andros Island and to the south and 
east by large areas of very shallow banks in the Bahamas.   This peculiar
geographical configuration, with shipping access only from the north through 
the narrow Providence Channel, provides quiet conditions, because of the 
low boat traffic and sluggish currents.

Individual hydrophones at Sites 3 and 4 are mounted at the ends of
4.5~m long booms extending above the ocean floor.  A total of 52
sensors cover the two sites, arranged on a triangular lattice with
2.5~km sides.  Analog signals from each hydrophone are preamplified
and brought to shore where digitizers and processors are located.  The
frequency response of the hydrophone and analog chain is flat to
within $\pm 5$~dB in the range from 1 to 50~kHz, while the sampling
rate of the digitizers is about 100~kHz.  Accurate GPS time stamping
is provided in the data stream.  In its normal operation the system is
capable to gather highly accurate 3-dimensional in-water tracking
data.  While a denser sensor spacing would of course be desirable, the
shower-to-hydrophone distance of 1~km we use though this work is close
to the worst case scenario of a neutrino interacting half-way between
two sensors.

\begin{figure}
\includegraphics{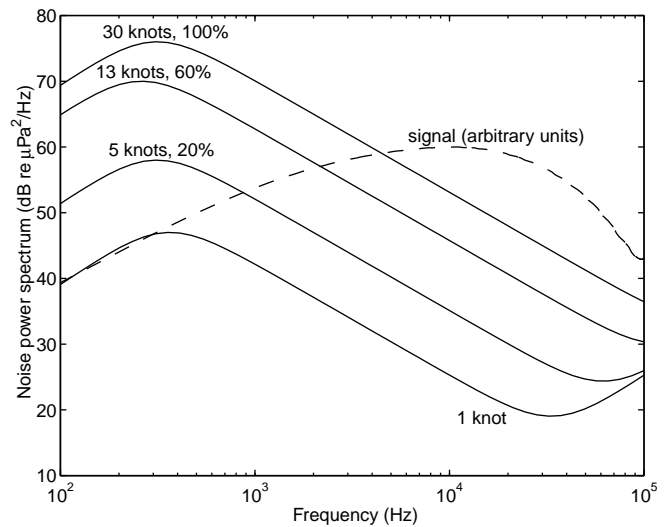}
\caption{Ambient noise spectra at AUTEC for different wind speeds 
(in knots, $1\ {\rm knot}=0.5144\ {\rm m/s}$) and the approximate
spectrum of the expected neutrino signal.  The absolute
amplitude of the signal is, of course, dependent upon the energy of
the primary and the location of the impact relative to the hydrophone.
The percent figure given next to each curve represents the cumulative
probability of finding such (or better) conditions at a given time.}
\label{fig:spectra}
\end{figure}

In Figure~\ref{fig:spectra} we show the random noise levels at AUTEC for
different wind conditions, along with the cumulative probability for
such conditions to occur on range.  The approximate frequency spectrum of 
the expected neutrino signals is also given for reference.    
The noise spectra in the Figure are due to the waves at the ocean surface 
induced by wind~\cite{wenz62,knudsen48,urick75} and, above $\gtrsim$10~kHz,
to thermal noise~\cite{wenz62}.   In addition coherent noise from 
human activities and natural phenomena should be considered.   A-priori we
expect two types of artificial noise: the first, due to ship screws, is mainly
confined to low frequency and hence easy to reject, while the second, due to
the range ``pingers'', is concentrated at a few well known frequencies 
and hence easy to filter-out.    In general exercises are performed on range
roughly 50\% of the time, so that very little man-made noise is expected
for a substantial fraction of every day (typically outside of working hours).
More serious is probably the high frequency noise produced by marine mammals 
using sonar to localize their prey and snapping shrimps~\cite{shrimp}.   
The severity of 
these backgrounds depends on the season and can only be quantitatively 
understood by analyzing a substantial amount of data from the array.
Here we limit ourselves to a detailed analysis of the random noise, 
formulating a filtering algorithm that can be implemented on the digital
signal processors (DSP) that analyze on-line the time-series from each
hydrophone.  This system of data selection can be used essentially as
a trigger signal to log an interval of the data stream for the entire
array (or maybe only some subset of sensors near to the one producing
the trigger).  Further data reduction, possibly involving the correlation 
of signals from different sensors, can then be done off-line.

Scattering of sound off the ocean surface and bottom may
alter somewhat the simple picture given above.    For instance we 
expect that each hydrophone will generally record two pulses: the direct
one and the one scattered from the ocean floor immediately around 
the sensor.  The time delay will be of the order of $h/c \sim 3$~ms, 
where $h=4.5$~m is the height of the hydrophones above the bottom. 
This effect can be used to estimate the angle of incidence of the 
acoustic wave and hence of the shower axis.      
In addition scattering phenomena can make the event detectable by more
than one hydrophone.    The attenuation and re-emission patterns from
scattering at the sea-bed and surface (where bubbles play an important
role) can be numerically estimated~\cite{beckmann63,crowther80}.
While large attenuations are to be expected so that it appears unlikely 
that these phenomena can be used to trigger an event that for geometrical
reasons was not directly visible, it is probable that scattered signals
can be detected {\it below} threshold by sensors in the vicinity of the
triggering hydrophone.   While off-line study of these correlations 
would be particularly important in confirming the event and in measuring 
position, orientation and energy of the shower, in the rest of this paper 
we concentrate on the triggering function that has to rely on single sensors.

\section{The sensitivity of the detector}
\label{sec:detection}

\begin{figure}
\includegraphics{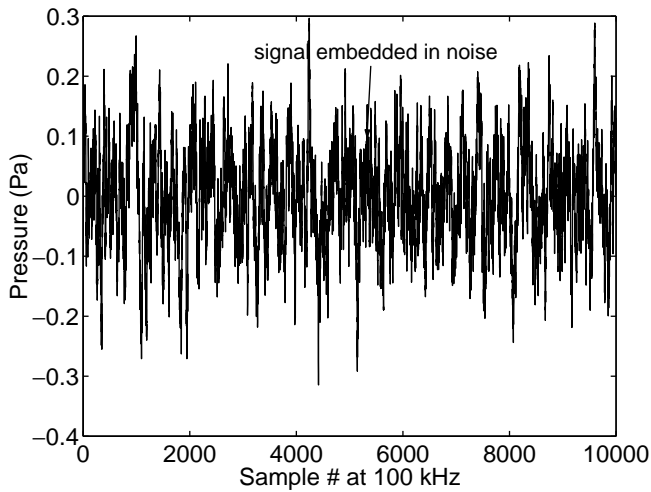}
\caption{
The signal from $\sci{2}{19}$~eV hadronic shower artificially embedded
in the noise sample of length $M=10000$ at position $k_0=5000$. The
noise is simulated using the spectrum for wind speed of 13 knots.
The time interval of this sample is 0.1~s.}
\label{fig:xdata}
\end{figure}

In order to analyze the problem of signal detection in some detail we
assume a discrete data sample set $x_k$ ($k=0,\dots,M-1$) with a
sampling rate of $f_s=100$~kHz, as provided by the AUTEC digitizers.
We consider the detection from a single site, as it
is natural given the small probability that the radiation pattern
discussed in the previous section would intercept more than one
hydrophone.   Sub-threshold use of signals from neighboring hydrophones
may be possible in off-line analysis but it is not discussed here. 
The data are the sum of the stationary Gaussian noise of a
given power spectrum and a signal of a given shape, but unknown
amplitude, starting at position $k_0$ in the data set, as illustrated
in the simulated time series of length $M=10000$ in
Figure~\ref{fig:xdata}. In this Figure a signal from a
$\sci{2}{19}$~eV hadronic shower at 1~km with the optimal angular
orientation to illuminate a hydrophone has been superimposed at a random
time (chosen in this case to start at sample number $k_0=5000$) over
the statistical noise spectrum relative to 13 knots wind from
Figure~\ref{fig:spectra}.     These conditions (or better) 
occur at the AUTEC site 60\% of the time.  The sampling frequency is high 
enough to provide negligible distortions.   
 
We extract the signal using the algorithm described in Appendix, which is
based on the digital filter of transfer function $H_l$ calculated on the 
basis of the signal shape and the Gaussian noise spectrum given in 
equation (\ref{eq:H}).    The variable $Y_k$ resulting from the 
application of the filter 
\[
Y_k=\sum_{l=k}^{k+N-1} H_l x_{k+l}
\]
has a Gaussian distribution and is used to assert the presence of the 
signal, using a threshold $Y_{\rm th}$ that can be chosen for a certain 
detection efficiency and false-alarm rate at a given signal amplitude $A$
($Y_k>Y_{\rm th}$ indicating the presence of a signal). In practice the
value of $N$ can be made small enough for efficient calculation on the
commercial DSP processors used at AUTEC.      

\begin{figure}
\includegraphics{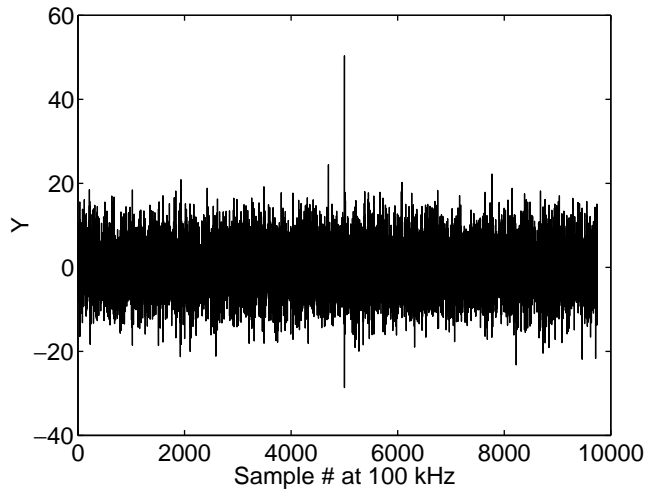}
\caption{The variable $Y$ discussed in the text and computed in detail 
in Appendix (see equation \ref{eq:Y}) for the time series shown in 
Figure~\ref{fig:xdata}.   A prominent peak at the correct location of
$k_0=5000$ is clearly visible.}
\label{fig:ysignal}
\end{figure}

In Figure~\ref{fig:ysignal} we plot the variable $Y_k$ for the
time series in Figure~\ref{fig:xdata}.  We see that $Y_k$ at the time
sample $k=k_0 = 5000$ has a value substantially higher that elsewhere.
Using a threshold $Y_{\rm th}=Y_{k_0}$, we find the probability of false 
alarm to be $\sci{5}{-16}$ at each data point. 
On the other hand the naive technique of trying to find the
signal using an amplitude threshold in time domain would give a false
alarm probability of 0.14 at each sample point, with a threshold set at 
amplitude of the signal.   The effectiveness of the method, also evident by
simple inspection of Figures~\ref{fig:xdata} and~\ref{fig:ysignal}, is 
based on the difference between the signal and noise spectra in
Figure~\ref{fig:spectra}.   This technique is equivalent to the application
of a matched filter~\cite{bradner77,cappellini78}. A value of $N=7$ was
used in the calculation, giving $\sim$15 floating-point operations
at each data point, which corresponds to a processor speed
requirement of $1.5$~Mflop/s.
The duration and amplitude of the signal can be estimated by maximizing a 
likelihood function.

We now use the probabilities of a signal miss and a false alarms found
in the Appendix to quantitatively analyze the power of the array as a 
UHE neutrino detector.   Here we stress the fact that our analysis does not
take into account possible interference from coherent sources, as already
mentioned above.

\begin{figure}
\includegraphics{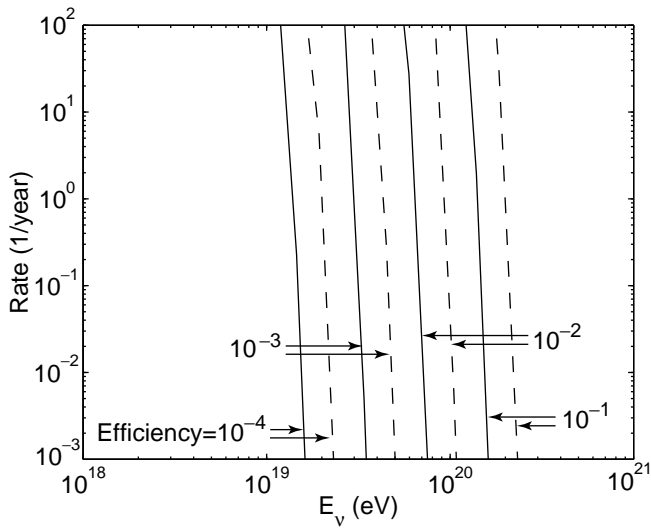}
\caption{
False alarm rates $\lambda$ for 
different detection efficiencies $\efficiency$.  
The solid 
lines are for $\nu_e$ or $\bar{\nu}_e$ (producing an electromagnetic
and an hadronic showers) while the dashed lines are for other neutrino 
flavors (where only the narrower beam from an hadronic shower is detected). 
The energy in abscissa is relative to the neutrino and it is shared between 
hadronic and leptonic components as expected for the energy of $10^{20}$~eV 
(20\% hadronic and 80\% leptonic). 
}
\label{fig:feasibility}
\end{figure}

We define the detector efficiency as the fraction of neutrinos that are 
detected respect to those interacting in the volume of water 
$\efficiency=\Lambda/\Lambda_0$.
We can write the interaction rate $\Lambda_0$ as
\[
\Lambda_0=F(E) N_A \rho_w \sigma(E) V_w
\]
where $F(E)$ is the UHE neutrino flux at energy $E$, $\sigma(E)$ is
the cross-section of interaction with a nucleon, $N_A\rho_w$ is the
number of nucleons in unit volume of water, and $V_w=ah$ is the volume
of the detector ($a$ and $h$ being the area and average depth of the
array). The detected rate is limited by the effective volume of the
radiation pattern $V_E$ in which the amplitude of the signal (relative
to produced by a shower of energy $10^{20}$~eV at a distance of 1~km)
is above the detection threshold $A$.  We can write $V_E$ as:
\begin{eqnarray}
\left(\frac{V_E}{1{\rm\ km}^3}\right)&=&
\frac{2\pi \Delta\theta}{3}\left(\frac{R_E}{1{\rm\ km}}\right)^3
\nonumber \\
&=&
\frac{2\pi \Delta\theta}{3}
\left[\left(\frac{E}{10^{20}{\rm\ eV}}\right)\frac{1}{A}\right]^3,
\label{eq:rpvolume}
\end{eqnarray}
where $\Delta\theta$ is the effective angular 
extent of the radiation lobe that can be found by integrating the curves
in Figures~\ref{fig:alvhad}b and~\ref{fig:lpm}b, and $R_E$ is its
radial extent.   We obtain 
$\Delta\theta\simeq 0.5^{\circ}$ ($\Delta\theta\simeq 1.5^{\circ}$) 
for showers induced by $\nu_\mu$,
$\bar{\nu}_\mu$, $\nu_\tau$, $\bar{\nu}_\tau$ ($\nu_e$,
$\bar{\nu}_e$).
The difference comes about from the electron neutrinos having the
electromagnetic part of the shower, while other flavors have only
hadronic part.
We note that the expression for $V_E$ includes a signal amplitude inversely
proportional to the distance $R_0$, as it should in our regime of 
radiation~\cite{learned79}.

The rate of detected neutrinos is then found by integrating over
volume:
\begin{equation}
\Lambda=N_h
F(E) N_A \rho_w  \sigma(E) 
\int_0^{V_{E,\rm max}}
p_{\rm detect}(A,Y_{\rm th})\,dV_E
\label{eq:flambda}
\end{equation}
where $N_h$ is the number of hydrophones and $p_{\rm detect}$ is the
detection probability given in (\ref{eq:pdetect}), with $A$ found from
$V_E$ using (\ref{eq:rpvolume}).
The upper
limit of the integration $V_{E,\rm max}$ is determined from the
condition that the acoustic radiation pattern is limited in size by
the size of the detector, i.e. $R_E<R_{\rm max}$, where $R_{\rm
max}\approx100$~km.
Finally, we get
\[
\efficiency=\frac{N_h}{V_w}
\int_0^{V_{E,\rm max}}
p_{\rm detect}(A,Y_{\rm th})\,dV_E
\]
We can now use this expression to set the value of $Y_{\rm th}$ for any
given efficiency $\efficiency$ and then calculate the resulting false
alarm rate as $\lambda=N_h f_s p_{\rm false}$, where $p_{\rm false}$
is determined from $Y_{\rm th}$ using (\ref{eq:pfalse}).
We plot the calculated false alarm rates for different efficiencies in
Figure~\ref{fig:feasibility}a. The solid lines correspond to $\nu_e$ or 
$\bar{\nu}_e$ that produce an electromagnetic and an hadronic showers, 
while the dashed lines are for other neutrino flavors that produce
a narrower sound lobe from the shorter hadronic shower.
As expected electromagnetic showers can afford a higher threshold that
results in a lower false alarm rate.

\section{Conclusions}
\label{sec:conclusions}

We have reviewed the possibility of searching for ultra-high-energy
neutrinos in cosmic rays using the acoustic emission from the
electromagnetic and hadronic showers produced by the neutrino
interactions in sea water.  We have analyzed the expected statistical
noise and devised an optimal algorithm to filter it out of the data
stream in real-time.  In a simulation we have applied this technique
to trigger the data-acquisition of a very large, high-frequency
multi-hydrophone array of the US Navy and found that one could trigger
on events at or above $\sim 10^{20}$~eV with tolerable false alarm
rates. Our algorithms are based on signals from individual sensors, as
appropriate for a trigger system, and are optimized to run on the
array's digital signal processors. This would make a data taking
campaign rather straightforward.  Further off-line analysis would be
needed to study the additional information that can be obtained by
multi-hydrophone correlations.  Although it is clear that an array
with higher sensor density would allow lower energy thresholds and
better redundancy, a test on the existent array analyzed here will
provide information on coherent noise and give a definite assessment
on the power of this technique.

\begin{acknowledgments}

We would like to thank D.~Bryant, J.~Cecil, N.~DiMarzio, T.~Kelly-Bissonnette, 
D.~Moretti (US Navy) for the many discussions on the neutrino detection 
capabilities of AUTEC.  We are also indebted to J.~Vandenbroucke (Stanford) 
for the help in programming the AUTEC signal processors. One of us (G.G.) is 
indebted to D.~Kapolka (US Navy) for early guidance in understanding the high 
frequency capabilities of different arrays.  We thank P.~Gorham (JPL) for a
critical reading of an early version of the manuscript.
This work was supported, in part, by a Terman Fellowship from the Stanford
University.  Partial support was also provided by the Office of Naval Research
under Grant No. N00014-93-1-0054.

\end{acknowledgments}

\appendix
\section{The detection algorithm}

In order to analyze the problem of signal detection in some detail we
assume a discretized data sample from a single hydrophone.  Since the
amplitude of the signal can vary with respect to the distance to the
source, we only fix the {\it shape} of the signal, without specifying
its amplitude in advance.  Let us assume that the signal shape after
discretization with sampling frequency $f_s$, is represented as an
array of N real numbers, $F_n$, where $n=0,\dots,N-1$, so that
the signal is $AF_n$, where $A$ is the amplitude.

The discrete data sample $x_k$, $k=0,\dots,M-1$, which is discretized
with sampling frequency $f_s$, is the sum of the stationary Gaussian
noise $w_k$ of a given power spectrum and a signal embedded at a
random position $k_0=0,\dots,M-N$ with an unknown amplitude $A$:
\[
x_k=\left\{
\begin{array}{ll}
w_k, & k<k_0 {\rm\ or\ } k\ge k_0+N \\
w_k+AF_{k-k_0}, & k_0\le k<k_0+N.
\end{array}
\right.
\]

We will search for the signal position $k_0$ and the amplitude $A$
which maximizes the conditional probability of a signal to be present
within a given data set. This is a regression problem that can be solved by
finding the maximum in the appropriate likelihood function~\cite{hogg95reg}. 
Let us take a subsample of the original data $X_k\equiv x_{k_0+k}$,
where $k=0,\dots,N-1$, and analogously define a noise subsample
$W_k\equiv w_{k_0+k}$. The likelihood of the signal presence at
position $k_0$ is given by the Bayes formula
\begin{eqnarray}
\lefteqn{\Prob{F,A\in[A,A+dA]|X}} \nonumber \\
& & {}=\frac{
\Prob{X|F,A} \Prob{F} f(A) dA
}{\Prob{X}}
\end{eqnarray}

The ratio to the probability of the absence of signal (which is found
in a similar manner) is
\begin{eqnarray*}
\lefteqn{\frac{\Prob{F,A\in[A,A+dA]|X}}{\Prob{{\rm no\ signal}|X}}} \\
& & {}=
\frac{\Prob{X|F,A} \Prob{F} f(A) dA}{\Prob{X|{\rm no\ signal}}
\Prob{{\rm no\ signal}}}
\end{eqnarray*}
We notice that some of these conditional probabilities are related to
the distribution of the Gaussian noise:
\begin{eqnarray*}
\Prob{X|F,A}&=&f_W(X-AF)\,dX \\
\Prob{X|{\rm no\ signal}}&=&f_W(X)\,dX
\end{eqnarray*}
where $dX=\prod_k dX_k$. The probability distribution $f_W(W)$ of the
Gaussian noise, defined so that $f_W(W)dW$ is the probability that
the noise values $W_k$ are in the interval $dW=\prod_{k=0}^{N-1}dW_k$,
is given by
\[
f_W(W)=C \exp\left(-\frac{1}{2}W^T K^{-1}W\right)
\]
where $K$ is the noise covariance matrix
defined as $K_{jk}=\langle W_j W_k \rangle$ that, for a stationary noise,
has the property of being a function of the difference $(k-j)$ only.
If this portion of the time series is long enough so that the noise values 
separated by $> N/2$ points are not correlated, we can assume circular
stationarity $K_{jk}= R_{(k-j)\mod N}$, where $R$ is an array of
length $N$.

The calculations are easiest in a basis in which the covariance
matrix is diagonal. For a circularly stationary signal, this is the
basis of Fourier components
\begin{equation}
\tilde{W}_k=\sum_{n=0}^{N-1} W_n \exp\left(-\frac{2\pi
i}{N}kn\right), \; k=0,\dots,N-1
\label{eq:fourier}
\end{equation}

The distribution of the noise Fourier components is defined so that
$f_{\tilde{W}}(\tilde{W})d\tilde{W}$ is the probability of $\tilde{W}$
to be in the interval
$d\tilde{W}=d\tilde{W}_0d\tilde{W}_{N/2}
\prod_{k=1}^{N/2-1}d[\Re{\tilde{W}_k}]d[\Im{\tilde{W}_k}]$.
Hence the probability distribution is given by
\[
f_{\tilde{W}}(\tilde{W})=\tilde{C} \exp\left(-\frac{1}{2}\sum_k |\tilde{W}_k|^2/S_k\right)
\]
where  $S_k$, $k=0,\dots,N-1$ is the noise spectrum defined as
$\langle\tilde{W}_k^*\tilde{W}_l\rangle=\delta_{kl}S_k$. This is related to the
covariance matrix as $S=N\tilde{R}$, with $R$ defined above.

Substituting $f_{\tilde{W}}(\tilde{W})$ we get the ratio of the likelihoods
\[
\frac{\Prob{F,A\in[A,A+dA]|X}}{\Prob{{\rm no\ signal}|X}}=
\frac{\Prob{F}}{\Prob{{\rm no\ signal}}} e^{L'} dA
\]
where the argument of the exponential function is
\[
L'=-\frac{1}{2}\sum_k\frac{1}{S_k}\left(
|\tilde{X}_k-A\tilde{F}_k|^2-|\tilde{X}_k|^2
\right)+\ln f(A)
\]
Instead of maximizing the original probability, we maximize the
log-likelihood $L'$.

Since $A$ is always positive and $\ln f(A)\approx \const$
for $A>0$, we choose the optimal amplitude estimate $\bar{A}$ which
maximizes $L'$:
\[
\bar{A}=\max\left\{0,\frac{\sum_k \Re(\tilde{X}_k\tilde{F}_k^*)/S_k
}{\sum_k|\tilde{F}_k|^2/S_k}
\right\}
\]
Substituting this value into $L'$, we form a random variable
\[
L=-\frac{1}{2}\sum_k\frac{1}{S_k}\left(
|\tilde{X}_k-\bar{A}\tilde{F}_k|^2-|\tilde{X}_k|^2
\right)=L'(\bar{A})+\const.
\]
Note that the maximization with respect to $A$ can be generalized to
other parameters of the signal. Let $F^{(m)}$ be different signal
shapes from a set given by an index $m$. Then to get the maximum
likelihood of the signal shape we choose the value of $m_0$ such that
$L^{(m_0)}=\max\{L^{(m)}\}$ (note that $\const=\ln f(A)$ is the same
for all $m$).

It turns out that the equation for $L$ can be simplified in terms of
the number of calculations required. Substituting into it the expression for
$\bar{A}$, we derive
\[
L=\left\{
\begin{array}{ll}
\frac{Y^2}{2\sigma_Y^2}, & Y>0 \\
0, & Y\le0
\end{array}
\right.
\]
where 
$Y=\sum_k \Re(\tilde{X}_k\tilde{F}_k^*)/S_k$
and $\sigma_Y^2=\sum_k |\tilde{F}_k|^2/S_k$ can be shown to be the
variance of $Y$.   Substituting the Fourier transform definition
(\ref{eq:fourier}) for $\tilde{X}$, we finally obtain
\begin{equation}
Y=\sum_{l=0}^{N-1} X_l H_l
\label{eq:Y}
\end{equation}
where
\begin{equation}
{H_l=\sum_k (\tilde{F}_k/S_k)e^{2\pi i kl/N}}.
\label{eq:H}
\end{equation}
The variable
$Y$ is thus the result of applying a digital filter with response
function $H$ or transfer function $\tilde{H}_k=N\tilde{F}_k/S_k$. The
amount of calculation can be reduced even further, due to a fact that
the quantity of $H_k$ that are significantly different from zero is
$N^*<N$. The number $N^*$ is estimated to be $\sim \tau_{\rm sig}f_s$
because $H_k$ has the time duration of the same order as the duration
of the signal, $\tau_{\rm sig}$. This last fact is confirmed by a
direct calculation. Thus, for typical values of $f_s=10^5$~Hz and $\tau_{\rm
sig}\simeq 10^{-4}$~s, we have $N^*\sim 10$. The number of floating-point 
operations (flops) required is then $\sim 10^6$ per second, which is 
feasible on modern DSP processors capable of more that 50~Mflops/s.

Signal detection can then be implemented at every hydrophone utilizing 
the fact that $Y$ has different distributions $f_{Y,A}(Y)$ in the cases
presence ($A\not=0$) and absence ($A=0$) of a signal.   We 
chose a threshold value $Y_{\rm th}$ and decide that the
signal is present (absent) when $Y\ge Y_{\rm th}$ ($Y<Y_{\rm th}$). 
Since the distributions $f_{Y,A}(Y)$ for the cases of signal presence 
and absence in general overlap, there will be the possibility of false 
alarms and signal misses. The threshold value has then to be chosen in
an appropriate manner.

We now estimate the probabilities of false alarms and misses for a certain 
threshold $Y_{\rm th}$ by estimating the probability distribution
$f_{Y,A}(Y)$ of the variable $Y$ in the presence of a signal of
amplitude $A$ (including the no-signal situation as a particular case
$A=0$). Since $Y$ is a linear combination of Gaussian variables, the
distribution sought is also Gaussian
\begin{equation}
{f_{Y,A}(Y)=\frac{1}{\sqrt{2\pi}\sigma_Y}
e^{-(Y-A\sigma_Y^2)^2/(2\sigma_Y^2)}}.
\label{eq:fYA}
\end{equation}

The probability of missing a signal is then
\begin{eqnarray}
p_{\rm miss}&=&\int_{-\infty}^{Y_{\rm th}} f_{Y,A}(Y)\,dY
\nonumber \\
&=&
1-\frac{1}{2}\erfc{\frac{Y_{\rm th}-A\sigma_Y^2}{\sqrt{2}\sigma_Y}},
\label{eq:pmiss}
\end{eqnarray}
while the probability of the signal detection is
\begin{eqnarray}
p_{\rm detect}=1-p_{\rm miss}&=&\int_{Y_{\rm th}}^\infty
f_{Y,A}(Y)\,dY
\nonumber \\
&=&
\frac{1}{2}\erfc{\frac{Y_{\rm th}-A\sigma_Y^2}{\sqrt{2}\sigma_Y}}.
\label{eq:pdetect}
\end{eqnarray}
Finally, the false alarm probability is
\begin{equation}
p_{\rm false}=\int_{Y_{\rm th}}^\infty  f_{Y,A=0}(Y)\,dL
=\frac{1}{2}\erfc{\frac{Y_{\rm th}}{\sqrt{2}\sigma_Y}}
\label{eq:pfalse}
\end{equation}
In the expressions above $\erfc{x}=1-\erf{x}$ is the complementary
error function.   The above expressions can be inverted to derive 
$Y_{\rm th}$ from $p_{\rm detect}$ and $p_{\rm false}$:
\begin{equation}
Y_{\rm th}=\left\{
\begin{array}{l}
A\sigma_Y^2+\sqrt{2}\sigma_Y \erfinv{1-2p_{\rm detect}} \\
\sqrt{2}\sigma_Y  \erfinv{1-2p_{\rm false}}
\end{array}
\right.
\label{eq:Ythp}
\end{equation}
where $\erfinv{x}$ is the inverse error function.  

\bibliography{draft}

\end{document}